\documentclass[12pt]{iopart}
\usepackage{iopams,graphicx}

\begin{document}

\title{On the terminal velocity of sedimenting particles in a flowing fluid}
\author{Marco Martins Afonso}
\address{Department~of~Mechanical~Engineering, Johns~Hopkins~University, 3400 North Charles Street, Baltimore,~MD~21218,~USA}
\ead{marcomar@fisica.unige.it}

\begin{abstract}
 The influence of an underlying carrier flow on the terminal velocity of sedimenting particles is investigated both analytically and numerically.
 Our theoretical framework works for a general class of (laminar or turbulent) velocity fields and, by means of an ordinary perturbation expansion
 at small Stokes number, leads to closed partial differential equations (PDE) whose solutions contain all relevant information on the sedimentation process.
 The set of PDE's are solved by means of direct numerical simulations for a class of 2D cellular flows (static and time dependent)
 and the resulting phenomenology is analysed and discussed.
\end{abstract}

\submitto{\JPA}

\maketitle

\section{Introduction}

 In many situations of interest, particles suspended in fluids cannot be modelled as simple point tracers.
 Both drops in gases and bubbles in liquids, and also solid powders in fluids,
 have a finite size and their density is, generally speaking, different from the one of the advecting fluid.
 The description of their movement must then take into account the effects of inertia: this is why such objects are usually called inertial particles.
 Generally speaking, understanding the dynamics of these impurities~\cite{BFF01,FP04} is very relevant in several domains,
 ranging from geophysics~\cite{FFS02,CFMS05,FSV06} to astrophysics~\cite{BCPS99,MM02}, and from industry to biology~\cite{KPSTT00,RMP04}.

 Here, our attention will be focused on the sedimentation of inertial particles seeded in a given flow and subject to the action of gravity.
 Our main aim is to obtain an Eulerian description of the sedimentation process starting from the well-founded Lagrangian viewpoint for particle
 motion in the limit of small inertia (\emph{i.e.}~when collision events can be neglected).
 Although our main focus is the sedimentation process, our theory provides the whole detailed statistical information of particle motion.
 The probability density function of having a particle in a given position at a certain time is indeed available from our approach.
 Our theoretical machinery, which applies for a wide class of velocity fields (either laminar or turbulent), is tested against numerical simulations
 (both direct numerical simulations and Lagrangian simulations) for a class of 2D cellular flows (static and time dependent).
 This specific example offers the possibility to analyse and discuss how the sedimentation process turns out to be extremely sensitive to the flow details.
 To be more specific, for the class of cellular flows, both an increase and a reduction of the rate of sedimentation with respect to the still-fluid case
 can occur, depending on the flow parameters like, \emph{e.g.},  flow geometry and pulsating frequency, as well as on the usual Stokes and P\'eclet numbers.
 Our findings lead to the following conclusion of interest in the realm of applications
 (\emph{e.g.}~within environmental sciences in connection to the problem of aerosol dispersion and sedimentation):
 any attempt to model the effect of flows on particle sedimentation cannot avoid taking into account full details of the flow field.

 The paper is organized as follows.
 In section~\ref{sec:ge} the basic equations governing the time evolution of inertial particles under gravity and in a prescribed flow are given,
 together with the associated Fokker--Planck equation for the particle probability density function.
 In section~\ref{sec:ssn} the problem of finding the terminal velocity of the sedimenting particles is formulated in a Hermitian form and tackled via a
 second-quantization formalism. The resulting set of auxiliary equations to obtain the terminal velocity are presented in the limit of small Stokes numbers.
 In section~\ref{sec:gf} the auxiliary equations are solved via direct numerical simulations in two dimensions and the obtained results are discussed.
 Conclusions and perspectives follow in section~\ref{sec:conc}. The appendix is devoted to some mathematical and computational details.

\section{General equations} \label{sec:ge}

 Let us consider the motion of a single, small, rigid, spherical particle of radius $b$ immersed in an incompressible $d$-dimensional
 flow $\bi{u}(\bi{x},t)$ (with $d\ge2$). Even if some of our results are more general, the flow field will be assumed
 either steady or periodic in time (with period $T$), and periodic in space (\emph{i.e.}~cellular, with period $L$)~\cite{MC86,M87a}.
 We shall focus our attention on the so-called Stokes regime, in
 which the surrounding flow is differentiable on scales of the order of $b$;
 we shall also neglect the feedback of the particle on the flow.
 The motion is thus influenced by gravity, buoyancy and drag~\cite{MR83}, to which Brownian
 noise should be added in order to take into account the thermal fluctuations of the fluid. Moreover, we neglect
 the Basset, Faxen, Oseen and Saffman corrections~\cite{MR83} and other possible effects due to rotationality, high
 relative velocity or non-sphericity~\cite{M90}.

 To write the equation for the particle trajectory $\bi{X}(t)$, it is customary to introduce the
 covelocity $\bi{V}\equiv\dot{\bi{X}}-\beta\bi{u}(\bi{X}(t),t)$. Here,
 $\beta\equiv3\rho_{\mathrm{f}}/(\rho_{\mathrm{f}}+2\rho_{\mathrm{p}})$ is an
 adimensional coefficient, where $\rho_{\mathrm{p}}$ and $\rho_{\mathrm{f}}$ are the
 densities of the particle and of the fluid, respectively. According to the ratio between the
 two densities, $\beta$ ranges from 0 ($\rho_{\mathrm{f}}\ll\rho_{\mathrm{p}}$: heavy particles,
 like drops in gases) to 3 ($\rho_{\mathrm{f}}\gg\rho_{\mathrm{p}}$: light particles, like bubbles
 in liquids), and becomes 1 when the two densities are equal (and inertial effects absent).
 Therefore, the covelocity differs from the slip velocity ($\dot{\bi{X}}-\bi{u}(\bi{X}(t),t)$)
 by a term $(1-\beta)\bi{u}(\bi{X}(t),t)$, which vanishes only for neutral particles.
 The concept of slip velocity is probably more familiar in connection with the study of relative acceleration.
 In our case, the use of covelocity is due to the fact that it strongly simplifies the equations of motion,
 properly keeping into account the added-mass effect without any need to introduce time derivatives of the external flow.
 However, $\bi{u}(\bi{x},t)$ being completely known, from the knowledge of $\bi{V}$ one can immediately find the real velocity $\dot{\bi{X}}$.

 The Stokes (response) time is defined as $\tau\equiv b^2/3\nu\beta$, with $\nu$ the kinematic viscosity.
 Denoting with $\bi{g}$, $\kappa$ and $\bfeta(t)$ the gravity acceleration,
 the particle diffusivity and the standard white noise, respectively, we have:
 \begin{equation} \label{sist}
  \left\{\begin{array}{l}
   \dot{\bi{X}}=\bi{V}+\beta\bi{u}(\bi{X}(t),t)\\
   \dot{\bi{V}}=\displaystyle\frac{(1-\beta)\bi{u}(\bi{X}(t),t)-\bi{V}}{\tau}+(1-\beta)\bi{g}+\frac{\sqrt{2\kappa}}{\tau}\bfeta(t)\;.
  \end{array}\right.
 \end{equation}

 The study can be carried on in the corresponding phase space $(\bi{x},\bi{v},t)$.
 Let us consider the probability that, at time $t$, the particle is at location
 $\bi{x}$ with a covelocity $\bi{v}$, and denote it by $p(\bi{x},\bi{v},t)$.
 This satisfies the Fokker--Planck equation associated to the stochastic differential equation (\ref{sist}):
 \begin{eqnarray} \label{efp}
  \mathcal{L}p\equiv&\left\{\frac{\partial}{\partial t}+\frac{\partial}{\partial x_{\mu}}\left[v_{\mu}+\beta u_{\mu}(\bi{x},t)\right]+\frac{\partial}{\partial v_{\mu}}\bigg[(1-\beta)g_{\mu}\right.&\\
  &\left.\left.+\frac{(1-\beta)u_{\mu}(\bi{x},t)-v_{\mu}}{\tau}\right]-\frac{\kappa}{\tau^2}\frac{\partial^2}{\partial v_{\mu}\partial v_{\mu}}\right\}p&=0\;.\nonumber
 \end{eqnarray}
 The linear operator $\mathcal{L}$ inside the curly brackets of (\ref{efp}) carries a factor $\kappa/\tau^2$ in
 front of the Laplacian with respect to the velocity, which is the highest-derivative term. Therefore, we may expect some
 singularity in the limits of vanishing diffusivity or large Stokes time.

 The terminal settling velocity of the inertial particle~\cite{M87b,ACHL02}
 is defined as the average particle velocity
 \begin{equation} \label{velcad}
  \bi{w}\equiv\!\int\!\rmd t\,\frac{1}{T}\!\int\!\rmd\bi{x}\!\int\!\rmd\bi{v}\,\left[\bi{v}+\beta\bi{u}(\bi{x},t)\right]p(\bi{x},\bi{v},t)\;,
 \end{equation}
 where the temporal integration is clearly omitted in the presence of steady flows.
 In general, such \emph{renormalized} terminal velocity can differ from the
 \emph{bare} terminal velocity, \emph{i.e.}~the particle settling velocity in a still
 fluid, whose value can easily be found to be
 \begin{equation} \label{btv}
  \bi{w}^*=(1-\beta)\bi{g}\tau\;.
 \end{equation}
 The difference turns out to be:
 \begin{equation} \label{dv}
  \Delta\bi{w}\equiv\bi{w}-\bi{w}^*=\!\int\!\rmd t\,\frac{1}{T}\!\int\!\rmd\bi{x}\!\int\!\rmd\bi{v}\,\bi{u}(\bi{x},t)p(\bi{x},\bi{v},t)\;.
 \end{equation}
 In what follows, we shall only take into account flows with zero mean
 ($\int\!\rmd t\int\!\rmd\bi{x}\,\bi{u}(\bi{x},t)=0$) and possessing odd
 parity with respect to reflections in the vertical direction.
 Such flows are relevant to analyse the vertical component $\Delta w$ of
 the terminal velocity discrepancy because no mean contribution is present,
 and every (eventual) nonzero result, originated from the component of $p$
 antisymmetric in the vertical coordinate, is to be interpreted as due to
 preferential concentration in areas of rising or falling fluid.

 Our plan is thus to solve (\ref{efp}), $\mathcal{L}p=0$, for the phase-space
 density and to plug the result in expression (\ref{dv}) to find $\Delta w$.

\section{Analytical investigation at small Stokes number} \label{sec:ssn}

 We shall first focus on situations in which the Stokes time is much smaller than the flow typical time scale, so that the
 particle adapts its own velocity to the one of the underlying fluid very rapidly.
 More precisely, denoting with $L$ and $U$ the typical length scale and velocity of the fluid
 respectively, in this section we shall only deal with small Stokes number, $\mathrm{St}\equiv\tau/(L/U)$. As we are
 taking into account also gravity and diffusivity (besides inertia), a thorough adimensionalization of the problem
 also requires the introduction of the Froude and P\'eclet numbers, defined as $\mathrm{Fr}\equiv U/\sqrt{gL}$ and
 $\mathrm{Pe}\equiv LU/\kappa$, respectively. As a first result, we see immediately that the vertical component of the
 bare terminal velocity (\ref{btv}), assumed as positive if pointing downwards and written in units of $U$, rewrites as
 \begin{equation} \label{vbtv}
  w^*=(1-\beta)\mathrm{St}\,\mathrm{Fr}^{-2}\;.
 \end{equation}

 Let us adimensionalize $\bi{x}$ with $L$, $t$ with $L/U$ and $\bi{u}$ with $U$, and let us denote the new variables with
 the same letters as before. Equation (\ref{efp}) can now be written in adimensional form after introducing an appropriate
 variable for the rescaled covelocity: the correct one turns out to be $\bi{y}\equiv\sqrt{\tau/2\kappa}\,\bi{v}$. With
 this notation, the operator acting upon the density $p(\bi{x},\bi{y},t)$ (appropriately normalized in the new
 variables) becomes:
 \begin{eqnarray} \label{astoc}
  \mathcal{L}&=&-\mathrm{St}^{-1}\left[\frac{\partial}{\partial y_{\mu}}y_{\mu}+\frac{1}{2}\frac{\partial^2}{\partial y_{\mu}\partial y_{\mu}}\right]+\mathrm{St}^{-1/2}\left[\sqrt{\frac{2}{\mathrm{Pe}}}\,y_{\mu}\frac{\partial}{\partial x_{\mu}}\right.\nonumber\\
  &&\left.+\sqrt{\frac{\mathrm{Pe}}{2}}\,(1-\beta)u_{\mu}(\bi{x},t)\frac{\partial}{\partial y_{\mu}}\right]+\mathrm{St}^0\left[\frac{\partial}{\partial t}\right.\nonumber\\
  &&\left.+\beta u_{\mu}(\bi{x},t)\frac{\partial}{\partial x_{\mu}}\right]+\mathrm{St}^{1/2}\left[\sqrt{\frac{\mathrm{Pe}}{2}}\,\frac{1-\beta}{\mathrm{Fr}^2}G_{\mu}\frac{\partial}{\partial y_{\mu}}\right]\nonumber\\
  &\equiv&-\mathrm{St}^{-1}\mathcal{A}_0+\mathrm{St}^{-1/2}\mathcal{A}_1+\mathrm{St}^0\mathcal{A}_2+\mathrm{St}^{1/2}\mathcal{A}_3\;,
 \end{eqnarray}
 where $\bi{G}\equiv\bi{g}/g$ is the unit vector pointing downwards, and the
 expressions of the operators $\mathcal{A}_i$ ($i=0,\ldots,3$) are easily
 identified from (\ref{astoc}). It is worth noticing that $\mathcal{A}_0$
 is an Ornstein-Uhlenbeck operator acting on covelocity coordinates only,
 which leads to a Hermitian reformulation of the problem in terms of a
 second-quantization formalism for the covelocity variable. Technical details
 on the analytical computation are left to the appendix: here we only recall
 the essential steps. By introducing a power-series expansion in $\mathrm{St}$
 for the particle probability density, we obtain a chain of advection--diffusion
 equations (see (\ref{equations})) for the auxiliary quantities
 $\psi_n^{\emptyset}(\bi{x},t)$ ($n\in\mathbb{N}$) defined in the appendix,
 in which lower-order quantities act as source terms for higher-order ones.
 Such equations can be solved sequentially, at least numerically, once the
 external flow is given. As a result, expression (\ref{dv}) for the
 variation of terminal velocity (in units of $U$) rewrites
 \begin{equation} \label{deldab}
  \Delta\bi{w}=\sum_{m=0}^{\infty}\mathrm{St}^{2+m}\!\int\!\rmd t\,\frac{L/U}{T}\!\int\!\rmd\bi{x}\,\bi{u}(\bi{x},t)\psi_{4+2m}^{\emptyset}(\bi{x},t)\;,
 \end{equation}
 where the integral on covelocity has already been performed (see (\ref{baru})). Note that only
 the vertically-antisymmetric components of $\psi_{4+2m}^{\emptyset}(\bi{x},t)$, hereafter
 denoted with the superscript $^{(\mathrm{a})}$ (in order to distinguish them from the
 vertically-symmetric counterparts $^{(\mathrm{s})}$), contribute to the spatial integral in
 (\ref{deldab}) for the vertical component $\Delta w$. Therefore, in order to study the leading
 contribution in $\Delta w$, quadratic in $\mathrm{St}$ (the equations needed to compute
 $\Or(\mathrm{St}^3)$ are not reported here), it is sufficient to consider the system
 \begin{equation} \label{equations}
  \left\{\begin{array}{l}
   \displaystyle\mathcal{M}\psi_4^{\emptyset(\mathrm{a})}=-(1-\beta)\mathrm{Fr}^{-2}G_{\mu}\frac{\partial}{\partial x_{\mu}}\psi_2^{\emptyset}\\
   \displaystyle\mathcal{M}\psi_2^{\emptyset}=(1-\beta)\frac{\partial u_{\mu}}{\partial x_{\nu}}\frac{\partial u_{\nu}}{\partial x_{\mu}}\psi_0^{\emptyset}\ \Longrightarrow\ \psi_2^{\emptyset}=\psi_2^{\emptyset(\mathrm{s})}\\
   \mathcal{M}\psi_0^{\emptyset}=0\ \Longrightarrow\ \psi_0^{\emptyset}=\textrm{const.}=\psi_0^{\emptyset(\mathrm{s})}\;,
  \end{array}\right.
 \end{equation}
 where we imposed uniform initial conditions (such that $\psi_0^{\emptyset}$ is a
 normalization constant) and denoted the adimensionalized advection--diffusion operator with
 \begin{equation} \label{oad}
  \mathcal{M}\equiv\frac{\partial}{\partial t}+u_{\mu}(\bi{x},t)\frac{\partial}{\partial x_{\mu}}-\mathrm{Pe}^{-1}\frac{\partial^2}{\partial x_{\mu}\partial x_{\mu}}\;.
 \end{equation}
 From (\ref{equations}), it is easy to understand that both $\psi_0^{\emptyset}$
 and $\psi_2^{\emptyset}$ do not depend on $\mathrm{Fr}$, and the former is
 clearly also independent of $\beta$. This means that $\psi_2^{\emptyset}$
 behaves like $(1-\beta)$, therefore $\psi_4^{\emptyset(\mathrm{a})}$
 is proportional to both $(1-\beta)^2$ and $\mathrm{Fr}^{-2}$.

 Let us now focus on this leading contribution at order $\mathrm{St^2}$ in $\Delta w$, corresponding to the
 term $m=0$ in (\ref{deldab}). From the analytical point of view, such considerations imply that we can introduce
 a non-universal (in the sense that it depends on $\mathrm{Pe}$ and possibly on other properties of the surrounding
 flow, \emph{e.g.}~its time frequency or space geometry, but not on $\mathrm{St}$, $\mathrm{Fr}$ or $\beta$) function
 \begin{equation} \label{dw}
  \Delta W\equiv\!\int\!\rmd t\,\frac{L/U}{T}\!\int\!\rmd\bi{x}\,G_{\mu}u_{\mu}(\bi{x},t)\frac{\psi_4^{\emptyset(\mathrm{a})}(\bi{x},t)}{\mathrm{Fr}^{-2}(1-\beta)^2}\;,
 \end{equation}
 such that we can write:
 \begin{equation} \label{main}
  \Delta w=[(1-\beta)\mathrm{St}/\mathrm{Fr}]^2\Delta W+\Or(\mathrm{St^3})\,.
 \end{equation}
 Equation (\ref{main}) summarizes the following two main analytical results for
 the variation of the actual terminal velocity in the limit of small inertia.
 i) It is influenced by gravity in the same way as the bare terminal velocity.
 Indeed, both $w^*$ in (\ref{vbtv}), and the coefficient in square brackets
 beside $\Delta W$ in (\ref{dw}), are proportional to $\mathrm{Fr}^{-2}$,
 \emph{i.e.}~to $\bi{g}$. This result is not trivial, because \emph{e.g.}~it
 does not hold, in general, at higher orders in $\mathrm{St}$.
 ii) It has the opposite parity with respect to the bare terminal velocity.
 Indeed, $w^*$ scales with $(1-\beta)$ while the above-mentioned coefficient
 with its square. In other words, if a specific flow under specific conditions
 is found to increase the falling of heavy particles, then the same flow under
 the same conditions (\emph{e.g.}~same $\mathrm{St}$, which may imply different $L/U$
 because $\tau$ depends on $\beta$) must simultaneously
 decrease the rising of light particles, and viceversa.

 The dependence on molecular diffusivity is more difficult to obtain analytically.
 In particular, as already pointed out, the large-$\mathrm{Pe}$ limit is singular,
 and nothing precise can be said about intermediate $\mathrm{Pe}$. On the contrary,
 for small $\mathrm{Pe}$, it can be shown rigorously that the behaviour of
 $\Delta W$ is $\Or(\mathrm{Pe}^2)$ or higher. As a particular case, we will show in
 the next section that, for the stationary square 2D Gollub flow, very well-defined
 asymptotics are found numerically for both small and large $\mathrm{Pe}$.
 In any case, one should keep in mind that ours is a series expansion in powers of $\mathrm{St}$,
 whose coefficients may become very large when varying the other parameters.
 In order for this to work, $\mathrm{St}$ should be assumed small enough
 so as to make the following terms negligible with respect to the previous ones.

\section{Numerical investigation: the cellular flow} \label{sec:gf}

 From the analytical point of view, the results reported in the previous section are the only available, and all refer
 to the $\Or(\mathrm{St}^2)$ correction. Proceeding at higher orders is not an easy task: \emph{e.g.}, the equation in
 (\ref{equations}) for $\psi_6^{\emptyset(\mathrm{a})}$ would be much more complicated, and involves as a
 source term also the vertically-symmetric component $\psi_4^{\emptyset(\mathrm{s})}$, whose equation is in turn by far
 more cumbersome than the one for the corresponding antisymmetric part. Clearly, this problem also reflects on the numerical
 approach: solving such equations by Direct Numerical Simulation (DNS) is beyond our possibilities. The DNS approach can
 however be applied to $\psi_2^{\emptyset}$ and to $\psi_4^{\emptyset(\mathrm{a})}$ if the surrounding flow is simple enough.

 An accurate study has been performed for the 2D Gollub flow, or, more precisely, for a simple generalization thereof:
 \begin{equation*}
  \left\{\begin{array}{l}
   u_1=\sin(kx_1)\cos[x_2+\sin(\omega t)]\\
   u_2=-k\cos(kx_1)\sin[x_2+\sin(\omega t)]\;,
  \end{array}\right.
 \end{equation*}
 where the parameters $k$ and $\omega$ represent respectively the cell aspect ratio (vertical
 extension over horizontal one) and the pulsation of the oscillations in the vertical direction
 $x_2$, assumed as positive downwards. The figures in this section all refer to the 2D Gollub
 flow in square cells ($k=1$) for heavy particles ($\beta=0$) at the specific value $\mathrm{Fr}=1$.

\begin{figure}
  \centering
  \includegraphics[height=4.5cm]{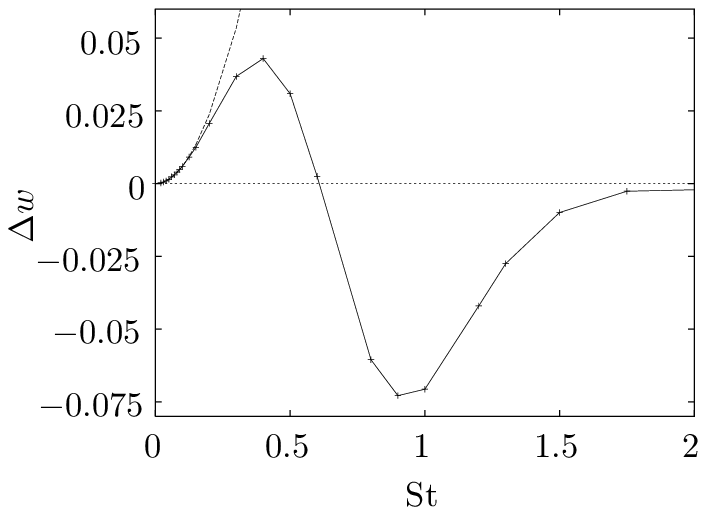}\hfill\includegraphics[height=4.5cm]{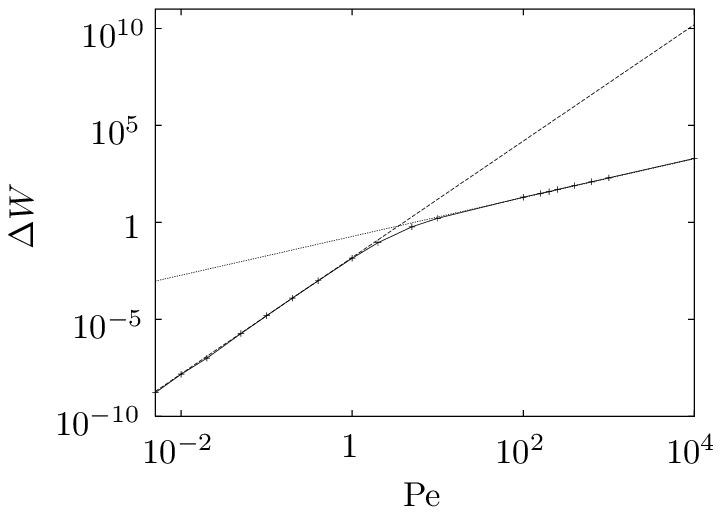}
  \caption{Left: comparison between LS (solid line) of (\ref{sist}) and the parabola (dashed line) resulting from the DNS solution of (\ref{equations}),
   for the variation of falling velocity as a function of the Stokes number at $\mathrm{Pe}=5$ and $\omega=0$.
   Right: dependence on the P\'eclet number for the variation of falling velocity (solid line) at $\Or(\mathrm{St}^2)$; the dashed line
   represents the cubic asymptote found analytically, while the dotted line is the linear asymptote with a multiplicative coefficient derived from a fit.}
  \label{fig1}
 \end{figure}

 First of all, we test our analytical result on the $\Or(\mathrm{St}^2)$ correction in the steady case
 ($\omega=0$). To do this, we performed a Lagrangian Simulation (LS) of the particle dynamics
 (\ref{sist}) at $\mathrm{Pe}=5$ (see, \emph{e.g.},~\cite{FMV98,FMV99,FMNV99}). Additionally, we solved the
 system (\ref{equations}) for $\psi_4^{\emptyset(\mathrm{a})}$ (and for $\psi_2^{\emptyset}$ in parallel)
 by means of DNS. In the left part of figure \ref{fig1},
 for small (and intermediate) values of $\mathrm{St}$, we report such LS and the parabola $\mathrm{St}^2\Delta W$, where
 $\Delta W$ is computed through (\ref{dw}) with the $\psi_4^{\emptyset(\mathrm{a})}$ found from DNS: the agreement is very
 good at least until $\mathrm{St}=0.15$. Moving at larger $\mathrm{St}$, it is evident how the increase in the falling velocity
 reaches a maximum value at $\mathrm{St}\simeq0.4$, then there is a crossover and a maximum reduction in settling takes place at
 $\mathrm{St}\simeq1$. The asymptotic vanishing at large $\mathrm{St}$ is a general feature in the ballistic limit, in which the
 particle reacts so slowly to the variations of the surrounding flow that the latter has basically no influence on settling.
 Such limit is singular, as already pointed out, because a particle with infinite inertia in a finite-correlated flow is
 equivalent to a particle with finite inertia in a $\delta$-correlated flow~\cite{BCH07a,BCH07b}, thus a technique like
 uniform-coloured-noise approximation should be used to deal with this situation properly.
 On the contrary, we think that our perturbative approach at small $\mathrm{St}$ should be able to capture at least the maximum in the plot,
 taking into account the following orders like $\psi_6^{\emptyset(\mathrm{a})}$ or higher.
 Of course, this behaviour could not be caught correctly by our previous quadratic approximation,
 which is not able to describe inflection points and subsequent changes in convexity.

 Let us now turn to the dependence of $\Delta W$ on $\mathrm{Pe}$. As already mentioned in the previous section, two
 well-defined asymptotics can be found, as shown in the right part of figure \ref{fig1}.
 In particular, at small $\mathrm{Pe}$, it can be shown by means of perturbative analysis that $\Delta W$ approaches
 $f(k)\,\mathrm{Pe}^3$, where $f(k)=(k^4+3k^2-2)/32(k^2+1)^2$ (\emph{i.e.}, $f(k)\ge0\Leftrightarrow k\ge(\sqrt{17}-3)/2\simeq0.56$).
 Note that this result is more stringent than the general behaviour $\Or(\mathrm{Pe}^2)$ (or higher) reported in
 the previous section, and only applies to this particular instance of flow. On the contrary, the linear
 asymptote $0.19\,\mathrm{Pe}$ at large $\mathrm{Pe}$ comes from a fitting procedure, because this limit is singular.
 Notice, however, that this does not necessary imply that the terminal velocity diverges at infinite $\mathrm{Pe}$:
 what we are discussing in this paragraph is only the behaviour of the $\Or(\mathrm{St}^2)$ term,
 and clearly there is the possibility of a regularizing resummation if one also considers higher-order terms.

\begin{figure}
  \centering
  \includegraphics[height=4.5cm]{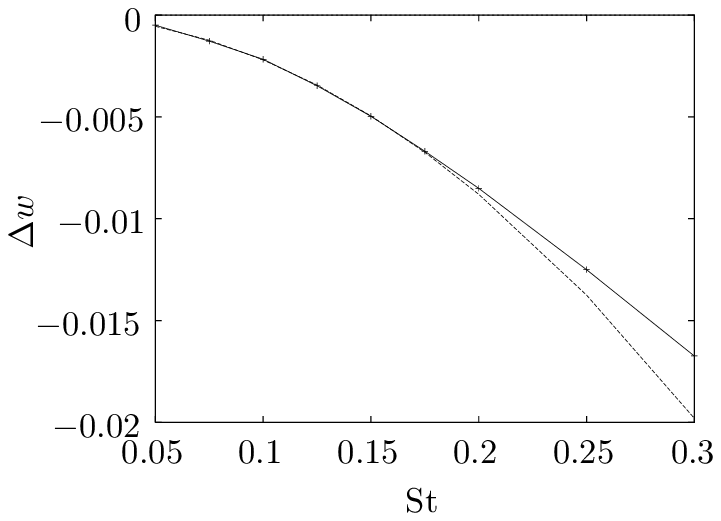}\hfill\includegraphics[height=4.5cm]{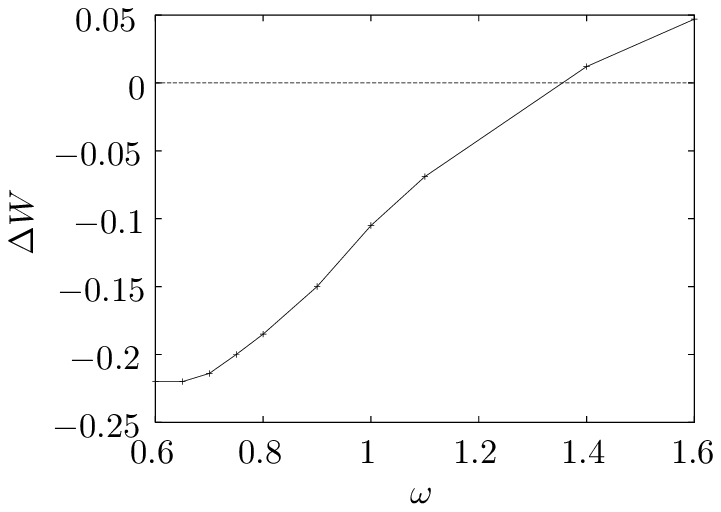}
  \caption{Left: same as in figure \ref{fig1} (left) but with $\omega=0.6$; notice the decrease in the terminal velocity opposed to the static-carrier-flow
   case. Right: dependence on the pulsation $\omega$ for the variation of falling velocity (solid line) at $\Or(\mathrm{St}^2)$, with $\mathrm{Pe}=5$.}
  \label{fig2}
 \end{figure}

 In figure \ref{fig2}
 we show that for a nonzero pulsating frequency the effect of the carrier flow may be the opposite (left panel).
 Indeed (right panel),
 the effective sedimenting velocity is decreased at small inertia in a window of values of $\omega$. For larger frequencies ($\omega>1.4$)
 an increase in the sedimenting velocity occurs again. This clearly shows how sensitive to the flow details the terminal velocity may be.

\section{Conclusions and perspectives} \label{sec:conc}

 We have investigated how the terminal velocity of sedimenting particles
 is influenced by a background flow. In the limit of small $\mathrm{St}$, we have
 reduced this problem to the solution of two coupled forced advection--diffusion
 equations. Such equations hold for generic velocities, laminar or turbulent,
 possessing odd parity with respect to reflections in the vertical direction.
 From our analysis the complete knowledge of the particle motion is available:
 the probability density function of having a particle in a given position at a
 given time is directly accessible once the differential equations are solved. The
 determination of the terminal velocity is a particular product of our analysis.
 Apart from the limiting case of small P\'eclet numbers, one has to resort to
 numerical simulations to find the latter. This task is easily accomplished for
 two-dimensional laminar flows, which clearly lead to the following conclusions.
 The value of $\mathrm{St}$ alone is not sufficient to argue if the sedimentation
 is faster or slower with respect to what happens in still fluid. In particular,
 for square, static cellular flows, at both intermediate and high $\mathrm{Pe}$,
 we found an increase of the falling velocity at relatively small $\mathrm{St}$,
 and a reduction starting from $\mathrm{St}$ larger than some critical value. A
 similar transition is also found for a fixed $\mathrm{St}$ and assuming the cell
 to be oscillating in the vertical direction with frequency $\omega$. We found the
 remarkable result that, appropriately tuning $\omega$, a decrease in settling can
 also be obtained at small $\mathrm{St}$, accompanied by an increase now occurring
 at sufficiently large $\mathrm{St}$. We also studied the effect of the flow spatial
 geometry on the terminal velocity. In the limit of small $\mathrm{Pe}$, the
 expression for the terminal velocity can be extracted analytically from our
 equations and its explicit dependence on the cell aspect ratio isolated. Changing the
 cell aspect ratio appropriately can cause a reduction in falling at small $\mathrm{St}$.

 It would be interesting to make a comparison between our method and the well-known
 ``continuum'' approximation~\cite{BFF01} to obtain the particle probability
 density. In the latter case, diffusivity is neglected from the beginning. Some
 preliminary results indicate that at the lowest order in $\mathrm{St}$ the correction
 to the bare falling velocity coincide when computed with the two strategies. Substantial
 differences start to emerge at higher orders in $\mathrm{St}$, where the ``continuum''
 approximation probably ceases to hold, due to the formation of caustic patterns
 and to the phenomenon of clustering.

\ack

 The author has been partially supported by COFIN 2005 project n.~2005027808.
 Antonio Celani and Andrea Mazzino are warmly acknowledged for their countless advice and support.

\appendix
\section{Mathematical and computational details}

 In this section we provide some details of the calculation which leads to (\ref{equations}).\\
 A Hermitian reformulation of the problem is convenient. Let us define
 $p_{\dag}(\bi{y})\equiv\exp(-y^2)$, the Gaussian kernel of $\mathcal{A}_0$:
 $\mathcal{A}_0p_{\dag}=0$. After decomposing the particle probability density as
 \begin{equation} \label{prima}
  p(\bi{x},\bi{y},t)=p_{\dag}^{1/2}(\bi{y})\psi(\bi{x},\bi{y},t)\;,
 \end{equation}
 our aim is to find the operators $\mathcal{B}_i$ such that
 $\mathcal{A}_ip=-p_{\dag}^{1/2}\mathcal{B}_i\psi$ ($i=0,\ldots,3$).\\
 This can be achieved by means of a second-quantization algorithm. One
 can indeed introduce the operators of creation and annihilation
 \begin{equation*}
  a_{\mu}^{\pm}=y_{\mu}\mp\frac{\partial}{\partial y_{\mu}}\;,
 \end{equation*}
 in terms of which
 \begin{equation*}
  y_{\mu}=\frac{1}{2}(a_{\mu}^++a_{\mu}^-)\;,\qquad\frac{\partial}{\partial y_{\mu}}=\frac{1}{2}(a_{\mu}^--a_{\mu}^+)
 \end{equation*}
 and the vacuum state turns out to be $a_{\mu}^-|0\rangle=0\Leftrightarrow|0\rangle=p_{\dag}^{1/2}$.\\
 Defining
 \begin{eqnarray*}
  \alpha_{\mu}=\sqrt{\frac{\mathrm{Pe}}{2}}(1-\beta)u_{\mu}(\bi{x},t)-\frac{1}{\sqrt{2\,\mathrm{Pe}}}\frac{\partial}{\partial x_{\mu}}\;,\\
  \gamma_{\mu}=-\frac{1}{\sqrt{2\,\mathrm{Pe}}}\frac{\partial}{\partial x_{\mu}}\;,\qquad\zeta_{\mu}=\sqrt{\frac{\mathrm{Pe}}{2}}\frac{1-\beta}{\mathrm{Fr}^2}G_{\mu}\;,
 \end{eqnarray*}
 the operators we are looking for read:
 \begin{eqnarray*}
  \mathcal{B}_0=\frac{1}{2}\left(y^2-\frac{\partial^2}{\partial y_{\mu}\partial y_{\mu}}-d\right)=\frac{1}{2}a_{\mu}^+a_{\mu}^-\;,\quad\mathcal{B}_3=\zeta_{\mu}a_{\mu}^+\;,\\
  \mathcal{B}_1=\alpha_{\mu}a_{\mu}^++\gamma_{\mu}a_{\mu}^-\;,\quad\mathcal{B}_2=-\frac{\partial}{\partial t}-\beta u_{\mu}(\bi{x},t)\frac{\partial}{\partial x_{\mu}}\;.
 \end{eqnarray*}
 Note that $\mathcal{B}_0$ can be interpreted as the occupation number, and the following commutation relations hold:
 \begin{equation*}
  [\mathcal{B}_0,a_{\mu}^{\pm}]=\pm a_{\mu}^{\pm}\;,\qquad[a_{\mu}^-,a_{\nu}^+]=2\delta_{\mu\nu}\;,\qquad[a_{\mu}^{\pm},a_{\nu}^{\pm}]=0\;.
 \end{equation*}
 On the other hand, $\mathcal{B}_2$ does not carry any creation or annihilation operator ($\mathcal{A}_2$ is
 independent of $\bi{y}$) and vanishes when under investigation are very heavy particles in a stationary state.

 The following step consists in the development
 \begin{equation} \label{seconda}
  \psi(\bi{x},\bi{y},t)=\sum_{n=0}^{\infty}\mathrm{St}^{n/2}\psi_n(\bi{x},\bi{y},t)
 \end{equation}
 and in the study of the set of equations arising at each (integer and half-i.) order in $\mathrm{St}$:
 \begin{equation} \label{step}
  \mathcal{B}_0\psi_n=\left\{\begin{array}{ll}
   0&\textrm{for }n=0\\
   \mathcal{B}_1\psi_{n-1}&\textrm{for }n=1\\
   \mathcal{B}_1\psi_{n-1}+\mathcal{B}_2\psi_{n-2}&\textrm{for }n=2\\
   \mathcal{B}_1\psi_{n-1}+\mathcal{B}_2\psi_{n-2}+\mathcal{B}_3\psi_{n-3}&\textrm{for }n\ge3\;.
  \end{array}\right.
 \end{equation}
 Such relations can be solved recursively by exploiting the simple inversion formula for a generic function $\Psi$:
 \begin{equation*}
  \mathcal{B}_0\Psi=a_{\mu_k}^+\cdots a_{\mu_1}^+|0\rangle\Rightarrow\Psi=\frac{1}{k}a_{\mu_k}^+\cdots a_{\mu_1}^+|0\rangle\;.
 \end{equation*}
 Therefore, one obtains
 \begin{eqnarray} \label{terza}
  \psi_n(\bi{x},\bi{y},t)=&\psi_n^{\emptyset}(\bi{x},t)|0\rangle+\psi_n^{\mu_1}(\bi{x},t)a_{\mu_1}^+|0\rangle+\ldots\nonumber\\
  &+\frac{1}{n!}\psi_n^{\mu_1\cdots\mu_n}(\bi{x},t)a_{\mu_n}^+\cdots a_{\mu_1}^+|0\rangle\;,
 \end{eqnarray}
 where
 \begin{eqnarray} \label{psin}
  \psi_n^{\mu_1\cdots\mu_k}(\bi{x},t)=\\
  \left\{\begin{array}{l}
   \alpha_{\mu_k}\psi_{n-1}^{\mu_1\cdots\mu_{k-1}}+\zeta_{\mu_k}\psi_{n-3}^{\mu_1\cdots\mu_{k-1}}+k^{-1}\mathcal{B}_2\psi_{n-2}^{\mu_1\cdots\mu_k}\\\ +2k^{-1}\gamma_{\mu_{k+1}}\langle\psi_{n-1}^{\mu_1\cdots\mu_{k+1}\cdots\mu_k}\rangle\hspace{0.5cm}\textrm{for }k=1,\ldots,n-2\\[0.2cm]
   \alpha_{\mu_k}\psi_{n-1}^{\mu_1\cdots\mu_{k-1}}\hfill\textrm{for }k=n-1,n
  \end{array}\right.\nonumber
 \end{eqnarray}
 and $\langle\ldots\rangle$ implies a symmetrization on the repeated index (\emph{i.e.}~the sum
 of the possible permutations of the repeated index, divided by the number of such terms).
 Note that, by definition, the quantity $a_{\mu_k}^+\cdots a_{\mu_1}^+|0\rangle$
 equals the multidimensional Hermite polynomial of degree $k$, $H_k^{\vec{\mu}}(\bi{y})$.

 At each step one must also impose the corresponding solvability
 condition, which forbids the presence of states proportional to
 $|0\rangle$ on every right-hand side of expressions (\ref{step}),
 in order to avoid inversion problems. These constraints give:
 \begin{equation} \label{condsol}
  2\gamma_{\mu}\psi_{n+1}^{\mu}+\mathcal{B}_2\psi_n^{\emptyset}=0\quad\forall n\in\mathbb{N}\;.
 \end{equation}
 By expressing each $\psi_{n+1}^{\mu}$, through (\ref{psin}), as a
 function of a combination of $\psi_m^{\emptyset}$ for some $m\le n$,
 expressions (\ref{condsol}) are to be interpreted as equations for the
 quantities $\psi_n^{\emptyset}$. Together with the normalization condition
 $\int\!\rmd\bi{x}\,\psi_n^{\emptyset}\propto\delta_{n0}$, they can be solved
 recursively (numerically, or even analytically in some fortunate circumstances)
 once the basic flow $\bi{u}(\bi{x},t)$ is given. All such equations are of the
 advection--diffusion type and are forced by lower-order, already-solved
 quantities, with the exception of the equations for $n=0$ and $n=1$, which are
 homogeneous and thus imply $\psi_0^{\emptyset}=\textrm{const.}$ and
 $\psi_1^{\emptyset}=0$ under our assumptions. Moreover, one should notice that
 even-order equations are forced only by even-order quantities, and the same happens
 with odd $n$'s: this means that $\psi_{2n+1}^{\emptyset}=0\ \forall n\in\mathbb{N}$;
 therefore, our expansion (\ref{seconda}) reduces to a series in integer powers of
 $\mathrm{St}$ only. In (\ref{equations}) we reported the first few equations
 corresponding to even $n$'s, separating the components symmetric and
 antisymmetric in the vertical direction. Note that the spatial gradient
 and the flow field are vertically antisymmetric, while the vertical unit
 vector $\bi{G}$ (reminiscent of gravity) is symmetric in this sense.

 Starting from the quantities $\psi_n^{\emptyset(\mathrm{a})}$ with even $n$, we can
 reconstruct the variation of the adimensionalized terminal velocity by substituting
 back (\ref{terza}) into (\ref{seconda}) into (\ref{prima}) into (\ref{dv}), and by making use
 of the orthonormalization of the Hermite polynomials $H_k^{\vec{\mu}}(\bi{y})$ in the integral
 in the $\bi{y}$ variable with weight $p_{\dag}^{1/2}(\bi{y})$ (note that $1=H_0^{\emptyset}$):
 \begin{eqnarray} \label{baru}
  \Delta\bi{w}&=\!\int\!\rmd t\,\frac{L/U}{T}\!\int\!\rmd\bi{x}\!\int\!\rmd\bi{y}\,\bi{u}(\bi{x},t)p(\bi{x},\bi{y},t)\nonumber\\
  &=\!\int\!\rmd t\,\frac{L/U}{T}\!\int\!\rmd\bi{x}\,\bi{u}(\bi{x},t)\!\int\!\rmd\bi{y}\,p_{\dag}^{1/2}(\bi{y})\psi(\bi{x},\bi{y},t)\nonumber\\
  &=\!\int\!\rmd t\,\frac{L/U}{T}\!\int\!\rmd\bi{x}\,\bi{u}(\bi{x},t)\!\int\!\rmd\bi{y}\,\mathrm{e}^{-y^2/2}\sum_{n=0}^{\infty}\mathrm{St}^{n/2}\psi_n(\bi{x},\bi{y},t)\nonumber\\
  &=\sum_{n=0}^{\infty}\mathrm{St}^{n/2}\!\int\!\rmd t\,\frac{L/U}{T}\!\int\!\rmd\bi{x}\,\bi{u}(\bi{x},t)\!\int\!\rmd\bi{y}\,\mathrm{e}^{-y^2/2}\times\nonumber\\
  &\qquad\times\sum_{k=0}^n\frac{1}{k!}\psi_n^{\mu_1\cdots\mu_k}(\bi{x},t)a_{\mu_k}^+\cdots a_{\mu_1}^+|0\rangle\nonumber\\
  &=\sum_{n=0}^{\infty}\mathrm{St}^{n/2}\sum_{k=0}^n\frac{1}{k!}\!\int\!\rmd t\,\frac{L/U}{T}\!\int\!\rmd\bi{x}\,\bi{u}(\bi{x},t)\psi_n^{\mu_1\cdots\mu_k}(\bi{x},t)\times\nonumber\\
  &\qquad\times\!\int\!\rmd\bi{y}\,\mathrm{e}^{-y^2/2}H_k^{\vec{\mu}}(\bi{y})\nonumber\\
  &=\sum_{n=0}^{\infty}\mathrm{St}^{n/2}\sum_{k=0}^n\frac{1}{k!}\!\int\!\rmd t\,\frac{L/U}{T}\!\int\!\rmd\bi{x}\,\bi{u}(\bi{x},t)\psi_n^{\mu_1\cdots\mu_k}(\bi{x},t)\delta_{k0}\nonumber\\
  &=\sum_{n=0}^{\infty}\mathrm{St}^{n/2}\!\int\!\rmd t\,\frac{L/U}{T}\!\int\!\rmd\bi{x}\,\bi{u}(\bi{x},t)\psi_n^{\emptyset}(\bi{x},t)\nonumber\\
  &=\sum_{m=0}^{\infty}\mathrm{St}^{m}\!\int\!\rmd t\,\frac{L/U}{T}\!\int\!\rmd\bi{x}\,\bi{u}(\bi{x},t)\psi_{2m}^{\emptyset}(\bi{x},t)\nonumber\\
  &=\sum_{m=0}^{\infty}\mathrm{St}^{m}\!\int\!\rmd t\,\frac{L/U}{T}\!\int\!\rmd\bi{x}\,\bi{u}(\bi{x},t)\psi_{2m}^{\emptyset(\mathrm{a})}(\bi{x},t)\nonumber\\
  &=\sum_{m=2}^{\infty}\mathrm{St}^{m}\!\int\!\rmd t\,\frac{L/U}{T}\!\int\!\rmd\bi{x}\,\bi{u}(\bi{x},t)\psi_{2m}^{\emptyset(\mathrm{a})}(\bi{x},t)\;,
 \end{eqnarray}
 which corresponds to (\ref{deldab}).


\Bibliography{20}
 \bibitem{BFF01}
  Balkovsky E, Falkovich G and Fouxon A 2001
  Intermittent distribution of inertial particles in turbulent flows
  {\it Phys. Rev. Lett.} {\bf 86} 2790--3
 \bibitem{FP04}
  Falkovich G and Pumir A 2004
  Intermittent distribution of heavy particles in a turbulent flow
  {\it Phys. Fluids} {\bf 16} L47--50
 \bibitem{FFS02}
  Falkovich G, Fouxon A and Stepanov M G 2002
  Acceleration of rain initiation by cloud turbulence
  {\it Nature} {\bf 419} 151--4
 \bibitem{CFMS05}
  Celani A, Falkovich G, Mazzino A and Seminara A 2005
  Droplet condensation in turbulent flows.
  {\it Europhys. Lett.} {\bf 70} 775--81
 \bibitem{FSV06}
  Falkovich G, Stepanov M and Vucelja M 2006
  Rain initiation time in turbulent warm clouds
  {\it J. Appl. Meteor. and Climatol.} {\bf 45} 591--9
 \bibitem{BCPS99}
  Bracco A, Chavanis P H, Provenzale A and Spiegel E A 1999
  Particle aggregation in a turbulent Keplerian flow
  {\it Phys. Fluids} {\bf 11}, 2280--7
 \bibitem{MM02}
  Matarrese S and Mohayee R 2002
  The growth of structure in the intergalactic medium
  {\it Mon. Not. R. Astr. Soc.} {\bf 329} 37--60
 \bibitem{KPSTT00}
  K\'arolyi G, P\'entek \'A, Scheuring I, T\'el T and Toroczkai Z 2000
  Chaotic flow: the physics of species coexistence
  {\it Proc. Natl. Acad. Sci.} {\bf 97} 13661--5
 \bibitem{RMP04}
  Ruiz J, Mac\'\i as D and Peters P 2004
  Turbulence increases the average settling velocity of phytoplankton cells
  {\it Proc. Natl. Acad. Sci.} {\bf 101} 17720--4
 \bibitem{MC86}
  Maxey M R and Corrsin S 1986
  Gravitational Settling of Aerosol Particles in Randomly Oriented Cellular Flow Fields
  {\it J. Atmos. Sci.} {\bf 43} 1112-–34
 \bibitem{M87a}
  Maxey M R 1987
  The motion of small spherical particles in a cellular flow field
  {\it Phys. Fluids} {\bf 30} 1915-–28
 \bibitem{MR83}
  Maxey M R and Riley J J 1983
  Equation of motion for a small rigid sphere in a nonuniform flow
  {\it Phys. Fluids} {\bf 26} 883-–9
 \bibitem{M90}
  Maxey M R 1990
  On the Advection of Spherical and Non-Spherical Particles in a Non-Uniform Flow
  {\it Phil. Trans. R. Soc. Lond.} A {\bf 333} 289-–307
 \bibitem{M87b}
  Maxey M R 1987
  The gravitational settling of aerosol particles in homogeneous turbulence and random flow fields
  {\it J. Fluid Mech.} {\bf 174} 441--65
 \bibitem{ACHL02}
  Aliseda A, Cartellier A, Hainaux F and Lasheras J C 2002
  Effect of preferential concentration on the settling velocity of heavy particles in homogeneous isotropic turbulence
  {\it J. Fluid Mech.} {\bf 468} 77--105
 \bibitem{FMV98}
  Frisch U, Mazzino A and Vergassola M 1998
  Intermittency in passive scalar advection
  {\it Phys. Rev. Lett.} {\bf 80} 5532--7
 \bibitem{FMV99}
  Frisch U, Mazzino A and Vergassola M 1999
  Lagrangian dynamics and high-order moments intermittency in passive scalar advection
  {\it Phys. Chem. Earth} {\bf 24} 945--51
 \bibitem{FMNV99}
  Frisch U, Mazzino A, Noullez A and Vergassola M 1999
  Lagrangian method for multiple correlations in passive scalar advection
  {\it Phys. Fluids} {\bf 11} 2178--86
 \bibitem{BCH07a}
  Bec J, Cencini M and Hillerbrand R 2007
  Heavy particles in incompressible flows: the large Stokes number asymptotics
  {\it Physica} D {\bf 226} 11-–22
 \bibitem{BCH07b}
  Bec J, Cencini M and Hillerbrand R 2007
  Clustering of heavy particles in random self-similar flow
  {\it Phys. Rev.} E {\bf 75} 025301(R)
\endbib
\end{document}